\documentclass[twocolumn]{aastex701}

\usepackage{soul}

\begin{document}

\title{JWST Reveals Refractory-Rich Water Ice in Interstellar Comet 3I/ATLAS: Evidence for a Continuum of Grain Properties across Protoplanetary Disks}

\author[0000-0001-8541-8550]{Silvia Protopapa}\affiliation{Solar System Science and Exploration Division, Southwest Research Institute, Boulder, CO 80302, USA}\email[show]{silvia.protopapa@swri.org}

\author[0000-0002-6702-7676]{Michael S. P. Kelley}
\affil{Department of Astronomy, University of Maryland, College Park, MD 20742-0001, USA}
\email{msk@astro.umd.edu}

\author[0000-0001-8233-2436]{Martin A. Cordiner}
\affiliation{Solar System Exploration Division, NASA Goddard Space Flight Center, 8800 Greenbelt Rd, Greenbelt, MD 20771, USA}
\affiliation{Department of Physics, Catholic University of America, Washington, DC 20064, USA.}
\email{martin.cordiner@nasa.gov}

\author[0000-0001-7694-4129]{Stefanie N. Milam}
\affiliation{Solar System Exploration Division, NASA Goddard Space Flight Center, 8800 Greenbelt Rd, Greenbelt, MD 20771, USA}
\email{stefanie.n.milam@nasa.gov}

\author[0000-0002-2668-7248]{Dennis Bodewits}
\affiliation{Physics Department, Edmund C. Leach Science Center, Auburn University, Auburn, AL 36849, USA}
\email{dennis@auburn.edu}

\author[0000-0001-6752-5109]{Steven B. Charnley}
\affiliation{Solar System Exploration Division, NASA Goddard Space Flight Center, 8800 Greenbelt Rd, Greenbelt, MD 20771, USA} \email{steven.b.charnley@nasa.gov}

\author[0000-0001-7479-4948]{Maria N. Drozdovskaya}
\affiliation{Department of Chemistry, Biochemistry and Pharmaceutical Sciences (DCBP), Universit{\"a}t Bern, Freiestrasse 3, 3012 Bern, Switzerland}
\email{maria.drozdovskaya.space@gmail.com}

\author[0000-0003-0194-5615]{Sara Faggi}
\affiliation{Department of Physics, American University, 4400 Massachusetts Ave NW, Washington, DC 20016, USA}
\affiliation{Solar System Exploration Division, NASA Goddard Space Flight Center, 8800 Greenbelt Rd, Greenbelt, MD 20771, USA}
\email{sara.faggi@nasa.gov}

\author[0000-0003-0774-884X]{Davide Farnocchia}
\affil{Jet Propulsion Laboratory, California Institute of Technology, 4800 Oak Grove Dr., Pasadena, CA 91109, USA}
\email[]{davide.farnocchia@jpl.nasa.gov}

\author[orcid=0000-0003-2354-0766]{Aurélie Guilbert-Lepoutre} 
\affiliation{LGL-TPE, CNRS, Université Lyon 1, ENSL, Villeurbanne, France}
\email{aguilbertlepoutre@gmail.com}

\author[0000-0001-7895-8209]{Marco Micheli}
\affil{ESA NEO Coordination Centre, Planetary Defence Ofvce, European Space Agency, Largo Galileo Galilei, 1, 00044 Frascati (RM), Italy}
\email[]{marco.bs.it@gmail.com}

\author[0000-0002-6006-9574]{Nathan X. Roth}
\affiliation{Solar System Exploration Division, NASA Goddard Space Flight Center, 8800 Greenbelt Rd, Greenbelt, MD 20771, USA}
\affiliation{Department of Physics, American University, 4400 Massachusetts Ave NW, Washington, DC 20016, USA}
\email{nathaniel.x.roth@nasa.gov}

\author[orcid=0000-0003-4365-1455]{Megan E. Schwamb}
\affiliation{Astrophysics Research Centre, School of Mathematics and Physics, Queen's University Belfast, Belfast BT7 1NN, UK}
\email[]{m.schwamb@qub.ac.uk} 

\author[0000-0002-0726-6480]{Darryl Z. Seligman}
\affiliation{Department of Physics and Astronomy, Michigan State University, East Lansing, MI, USA}
\email[]{dzs@msu.edu} 

\author[0000-0002-2662-5776]{Geronimo L. Villanueva}
\affiliation{Solar System Exploration Division, NASA Goddard Space Flight Center, 8800 Greenbelt Rd, Greenbelt, MD 20771, USA}
\email[]{geronimo.l.villanueva@nasa.gov}




\begin{abstract}
We present JWST/NIRSpec PRISM observations of the interstellar comet 3I/ATLAS obtained on 2025 August 6 (Epoch~1), 2025 December 22 (Epoch~2), and 2026 April 1 (Epoch~3), spanning eight months around perihelion at heliocentric distances of 3.3, 2.4, and 5.7~au, respectively. The spectra reveal a broad 3~\textmu{}m absorption band together with H$_2$O, CO$_2$, and CO gas emission. Unlike previously reported water-ice-rich Solar System comae, the strong 3~\textmu{}m absorption is accompanied by weak or absent 1.5 and 2.0~\textmu{}m water-ice bands. Spectral modeling indicates that the observations are best reproduced by submicron- to micron-sized water-ice-bearing aggregates containing refractory material. Compared with Epoch~1, the Epoch~3 spectrum favors the presence of a second population of larger, micron-sized, ice-rich aggregates and exhibits a subtle Fresnel-like structure near 3.1~\textmu{}m, consistent with crystalline water ice. The observations can be explained by either crystalline water ice at both epochs, with the spectral evolution arising primarily from changes in grain size and refractory mixing, or an evolution from an amorphous-like to crystalline state. The spectral properties of 3I bridge those of water-ice-bearing Solar System comae and several spectral classes of mid-sized trans-Neptunian objects, suggesting that the icy building blocks of planetesimals formed in different protoplanetary disks may span a continuum in the physical state of water ice, ranging from pure ice grains to water-ice-bearing aggregates with varying refractory content at submicron-to-micron scales, with 3I extending toward the refractory-rich end of this continuum.
\end{abstract}


\section{Introduction} 

Interstellar objects (ISOs) are commonly interpreted as planetesimals formed in extrasolar protoplanetary disks and subsequently ejected through gravitational interactions during planetary system assembly and evolution \citep{Dones2004,Levison2010,JewittSeligman2023}. As such, ISOs provide rare samples of solid material originating in extrasolar protoplanetary disks, offering a unique opportunity to constrain planetesimal formation, volatile retention, and disk chemistry beyond the Solar System. Comparative studies between ISOs and small-body populations in our own system---particularly comets, Centaurs, and small, undifferentiated trans-Neptunian objects (TNOs)---enable assessment of chemical and physical similarities among planet-forming environments.

The discoveries of 1I/`Oumuamua (1I), 2I/Borisov (2I), and, most recently, 3I/ATLAS (3I) have transformed ISOs from theoretical predictions into an observationally accessible population. In contrast to 1I, which exhibited no detectable coma \citep{Meech2017,Jewitt2017ApJ_1I}, cometary activity has been detected in both 2I and 3I \citep{Cordiner2020,Bodewits2020,Seligman2025ApJ,Cordiner2025}, allowing direct spectroscopic characterization of volatile and solid-state species released from their nuclei.

Water ice is a primary volatile constituent in cometary nuclei and on the surfaces of TNOs, and serves as a sensitive tracer of formation and evolutionary history \citep{Weissman2020SSRv,Brown2012}. Its near-infrared spectral features---most prominently the absorptions near 1.5, 2.0, and 3.0--3.2~\textmu{}m and their relative strengths---encode information on grain size, crystallinity, temperature, and the degree of mixing with refractory and other volatile species \citep{Mastrapa2008,Mastrapa2009,Davies1997,Protopapa2014}. In cometary comae, water-ice grains are often interpreted as recently ejected material that may retain signatures of the nucleus interior and, by extension, of the physical and chemical conditions prevailing in the protoplanetary disk midplane where planetesimals formed \citep{Ahearn2011,Protopapa2014,Protopapa2018,Sunshine2021}. However, such interpretations require careful consideration of the thermal evolution of grains after release from the nucleus \citep{Protopapa2021_46P,Kelley2025}. Water-ice grains are most readily studied at large heliocentric distances, typically beyond several au, where sublimation is less efficient and grain lifetimes are longer. However, comets in this regime are often too faint for detailed infrared spectroscopy, particularly beyond 2.5~\textmu{}m where water ice displays its strong fundamental O--H stretching absorption band. The sensitivity of JWST now enables spectroscopic characterization of icy grains across the 1--5~\textmu{}m wavelength range in faint and distant comets, extending such studies into regimes previously inaccessible from the ground.

3I provides a particularly compelling opportunity to investigate the properties of icy grains in an interstellar comet. Ground-based observations yielded conflicting evidence for water ice in the coma of 3I. Some studies reported evidence for a weak 2.0~\textmu{}m absorption feature interpreted as water ice \citep{Yang2025_3I}, whereas others found no obvious evidence for water-ice absorption bands in the near-infrared spectrum \citep{Kareta2025,Medler2026}. Subsequent JWST/NIRSpec observations revealed a coma dominated by CO$_2$ gas at the time of those observations, together with detections of H$_2$O gas, CO gas, and solid-phase water ice \citep{Cordiner2025}. The water-ice identification is supported by a strong absorption feature near 3~\textmu{}m, associated with the fundamental O--H stretching mode. However, unlike previously characterized water-ice-bearing cometary comae, the accompanying 1.5 and 2.0~\textmu{}m absorption bands are weak or absent in 3I. 

In this work, we present a detailed compositional and physical analysis of water-ice grains in the coma of 3I using JWST/NIRSpec spectroscopy obtained before and after perihelion. We focus on the morphology and relative strengths of the 1.5, 2.0, and 3~\textmu{}m water-ice absorption bands, together with the associated scattering continuum, to constrain grain size, crystallinity, and mixing with refractory material. Observations obtained near perihelion are used to place these results in the broader context of the comet's temporal evolution and activity. Finally, we compare the inferred ice properties with water-ice detections in Solar System comets and TNOs to assess whether the grains observed in 3I are consistent with known icy reservoirs and whether common physical processes govern the spectral appearance of water ice in icy bodies formed in different planetary systems.

\begin{figure*}[ht!]
\plotone{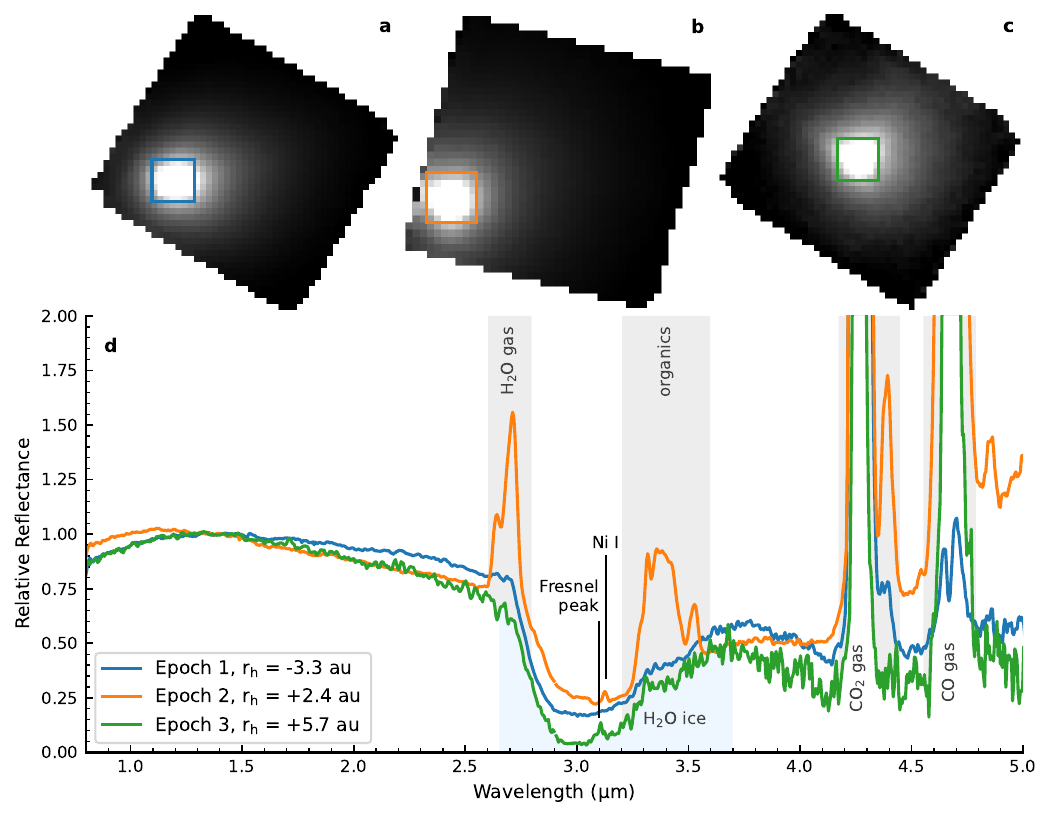}
\caption{
JWST/NIRSpec PRISM observations of 3I across three epochs. (a--c) Median-collapsed images of the spectral cubes from Epochs~1, 2, and 3, corresponding to JWST observation identifiers o001, o013, and o003, respectively, constructed from channels~10--267 over the 0.6--1.9~\textmu{}m wavelength range. Colored boxes indicate the $7\times7$ pixel extraction apertures used for the spectral analysis. The apparent offset of the comet from the center of the field of view results from a combination of ephemeris uncertainties and telescope pointing errors. (d) Extracted spectra from the three epochs, normalized over the 1.4--1.42~\textmu{}m interval. The spectra display the scattered solar continuum together with prominent spectral features, including H$_2$O gas emission near 2.7~\textmu{}m, the broad H$_2$O-ice absorption with a minimum near 3~\textmu{}m, organic-related spectral structures, and CO$_2$ and CO gas emission near 4.25 and 4.67~\textmu{}m, respectively. Gray shaded regions identify wavelength intervals affected by gas and organic features, while the blue shaded region marks the wavelength range associated with the broad H$_2$O-ice absorption inferred from the spectral modeling. The evolution of the continuum shape and spectral-feature strengths reflects changes in the coma composition and activity with heliocentric distance. \label{fig:general}}
\end{figure*}

\begin{deluxetable*}{ccccccccc}
\tablenum{1}
\tablecaption{Summary of the JWST/NIRSpec PRISM observations of 3I (Program \#5094).\label{tab:observations}}
\tablewidth{0pt}
\tablehead{
\colhead{Epoch} &
\colhead{Obs. ID} &
\colhead{Date} &
\colhead{Mid-time (UTC)} &
\colhead{$r_{\mathrm{h}}$ (au)\tablenotemark{a}} &
\colhead{$\Delta$ (au)\tablenotemark{b}} &
\colhead{$\alpha$ ($^\circ$)\tablenotemark{c}} &
\colhead{$T_{\mathrm{mag}}$\tablenotemark{d}} &
\colhead{Exp. Time / Dither (s)}
}
\startdata
1 & 001 & 2025-08-06  & 11:11:19 & -3.31 & 2.73 & 16.08 & 17.0 & 146 \\
2 & 013 & 2025-12-22  & 07:13:56 & 2.38 & 1.80 & 22.58 & 15.5 & 146 \\
3 & 003 & 2026-04-01  & 23:02:56 & 5.65 & 5.59 & 10.26 & 19.6 & 875 \\
\enddata
\tablenotetext{a}{Heliocentric distance; negative values indicate pre-perihelion observations.}\tablenotetext{b}{Observer--target distance.}\tablenotetext{c}{Sun--target--observer phase angle.}\tablenotetext{d}{Predicted apparent visual magnitude of the target used for JWST observation planning as reported by Horizons.}
\end{deluxetable*}
\section{Observations and Data Reduction}\label{sec:observations} 

3I was observed with JWST/NIRSpec as part of GO program 5094 (PI: Cordiner). Spectra were acquired with the NIRSpec PRISM mode \citep{Boeker2023} at three epochs: 2025 August 6 (Epoch~1), 2025 December 22 (Epoch~2), and 2026 April 1 (Epoch~3), spanning approximately eight months around perihelion on 2025 October 29 (Table~\ref{tab:observations}). The PRISM mode provides continuous spectral coverage from 0.6 to 5.3~\textmu{}m at low spectral resolution ($R\sim30$--300), enabling simultaneous characterization of the scattered solar continuum, broad solid-state absorption features, and gas emission \citep{Cordiner2025}.

Each epoch consisted of four dithered exposures. The effective exposure time per dither was $\sim$146~s for Epochs~1 and 2 and $\sim$875~s for Epoch~3, with the longer integration in Epoch~3 compensating for the expected decrease in target brightness (Table~\ref{tab:observations}). Because in-scene background subtraction is not feasible for an extended coma, four dedicated background exposures were obtained for each epoch, offset sufficiently far from the target position to avoid coma contamination ($180^{\prime\prime}$ for Epochs~1 and 2 and $300^{\prime\prime}$ for Epoch~3).

The uncalibrated data were retrieved from MAST and processed locally using Stage~1 of the JWST calibration pipeline \citep[v1.20.2,][]{bushouse_2024_10870758} with CRDS context \texttt{jwst\_1464.pmap}. For each epoch, the four dedicated background exposures were median-combined to construct a master background frame. Stage~2 processing was applied to each science dither individually, with \texttt{NSClean} \citep{Rauscher2024} and background subtraction enabled.

Spectra were extracted from each dither using an empirical profile-fitting technique employing wavelength-dependent source profiles derived directly from the target data within a $7\times7$ pixel extraction aperture, resulting in the highest signal-to-noise ratio among the extraction methods explored \citep{Wong2024,Protopapa2024,Protopapa2025}. The four dither spectra were averaged and corrected for flux losses using a wavelength-dependent aperture correction derived by comparing observations of the standard star SNAP-2 from program 01128, obtained with the same instrumental configuration and extraction methodology as the science data, to the corresponding CALSPEC model spectrum following \citet{Protopapa2024,Protopapa2025}.

The resulting flux-calibrated spectra (Fig.~\ref{fig:general}) were converted to relative reflectance using a PSG solar reference spectrum combining the Kurucz solar model with ACE-FTS solar measurements \citep{Villanueva2018}.
\section{Spectral modeling} \label{sec:modeling}
\subsection{Observed spectral properties} 
As shown in Fig.~\ref{fig:general}, the 3I spectra exhibit a blue near-infrared continuum together with a strong 3~\textmu{}m absorption band, while lacking prominent 1.5 and 2.0~\textmu{}m water-ice absorptions. The continuum slopes, measured over 1.15--2.50~\textmu{}m after normalization between 2.3 and 2.4~\textmu{}m, are $-1.1$ and $-2.0\%/100$~nm for Epochs~1 and 3, respectively. The corresponding 3~\textmu{}m band depths, measured over the 2.96--3.00~\textmu{}m interval relative to a linear continuum defined over 2.2--2.3 and 3.75--3.85~\textmu{}m, are 77\% and 94\%, respectively. Unlike Epochs~1 and 2, the Epoch~3 spectrum exhibits a subtle Fresnel-like structure near 3.10~\textmu{}m (Fig.~\ref{fig:general}), close to the expected position of the water-ice Fresnel reflection feature. This feature arises from anomalous dispersion in the refractive index within the broad O--H stretching band, and its morphology is more pronounced and structured for crystalline than for amorphous water ice (Appendix~\ref{sec:appendix_fresnel}, Fig.~\ref{fig:optical_constants}). Because the flux approaches zero near the center of the 3~\textmu{}m absorption band, we tested the robustness of this structure against individual dithers and different extraction methodologies (Appendix~\ref{sec:appendix_fresnel}, Fig.~\ref{fig:fresnel_robustness}). The broader morphology near 3.1~\textmu{}m is preserved across different
dither combinations and is also present in the independent aperture
extraction, although the detailed shape of the local feature is less robust. Epoch~2 instead displays a peak shifted toward longer wavelengths near 3.13~\textmu{}m, attributed to Ni~I emission in contemporaneously obtained higher-resolution observations \citep{Roth2026}. Collectively, these spectral differences indicate temporal evolution in the
properties of the coma ice population, including particle size, aggregate structure, refractory mixing, and ice
phase, as investigated through the modeling and synthetic calculations presented
below and in Appendix~\ref{sec:appendix_modeling}. 
\subsection{Modeling Methodology} 
We modeled only Epochs~1 and 3, as Epoch~2 exhibits substantially stronger contamination from H$_2$O gas, organic emission, and Ni~I emission features across the 3~\textmu{}m water-ice absorption region. In addition, unlike Epochs~1 and 3, Epoch~2 displays a long-wavelength excess beyond 4.2~\textmu{}m that we attribute to thermal emission from coma dust. The earlier onset of the thermal component in Epoch~2 increases the likelihood of overlap between scattered and thermal contributions within the 3~\textmu{}m region, further complicating the characterization of water ice \citep{Protopapa2014}.

The spectral modeling was restricted to the 1.15--4.2~\textmu{}m interval, which encompasses the principal water-ice absorption bands while avoiding regions dominated by strong CO and CO$_2$ gas emission. The 2.6--2.8 and 3.2--3.6~\textmu{}m spectral intervals, affected by weaker H$_2$O and organic emission features (Fig.~\ref{fig:general}), were retained in the fit but assigned inflated uncertainties to reduce their influence on the optimization. Because Epochs~1 and 3 do not exhibit evidence for a significant thermal contribution within the 1.15--4.2~\textmu{}m wavelength range, we modeled these spectra using only the scattered-light component of the coma reflectance.

Following previous analyses of cometary water-ice grains
\citep[e.g.,][]{Protopapa2018,Protopapa2021_46P},
we modeled the coma reflectance as an areal mixture of two
independent particle populations whose compositions ranged from pure
materials to aggregate grains composed of water ice and refractory
components. For each population, the equivalent single-scattering
albedo was computed by integrating Mie scattering and extinction
efficiencies over a differential particle-size distribution, $n(r)\propto r^{-\alpha}$, between the minimum and maximum particle radii, $r_{\min}$ and $r_{\max}$ (details are given in Appendix~\ref{sec:appendix_modeling}). Aggregate grains were represented using the Bruggeman
effective-medium approximation \citep{BohrenHuffman1983},
which approximates sub-particle-scale mixing among the constituent materials while preserving their individual optical properties. The observed reflectance was modeled as
\begin{equation}
R(\lambda)=F_{A}R_{A}+(1-F_{A})R_{B},
\end{equation}
where $F_A$ is the areal fraction of population~A, and
$R_A$ and $R_B$ are the reflectances of the individual
populations, approximated using the semi-infinite
diffuse-reflectance formulation \citep{Hapke2012}. Unlike traditional surface-oriented Hapke slab models, the
single-scattering albedo in the present implementation is
derived from Mie-computed scattering properties integrated over
a coma particle-size distribution, yielding a formulation more
representative of an optically thin coma composed of discrete
aggregate grains than of a semi-infinite particulate surface (Appendix~\ref{sec:appendix_modeling}).

The free parameters in the model include the areal fraction, \(F_A\), and, for each particle population, the minimum particle size, \(r_{\min}\), and differential particle-size distribution exponent, \(\alpha\). For aggregate populations, the relative volume fractions of the constituent components within each aggregate, \(\phi\), were treated as additional free parameters. Model parameters were constrained using a hybrid optimization approach combining deterministic fitting with Bayesian posterior sampling (Appendix~\ref{sec:appendix_modeling}).

\begin{figure*}[ht!]
\plotone{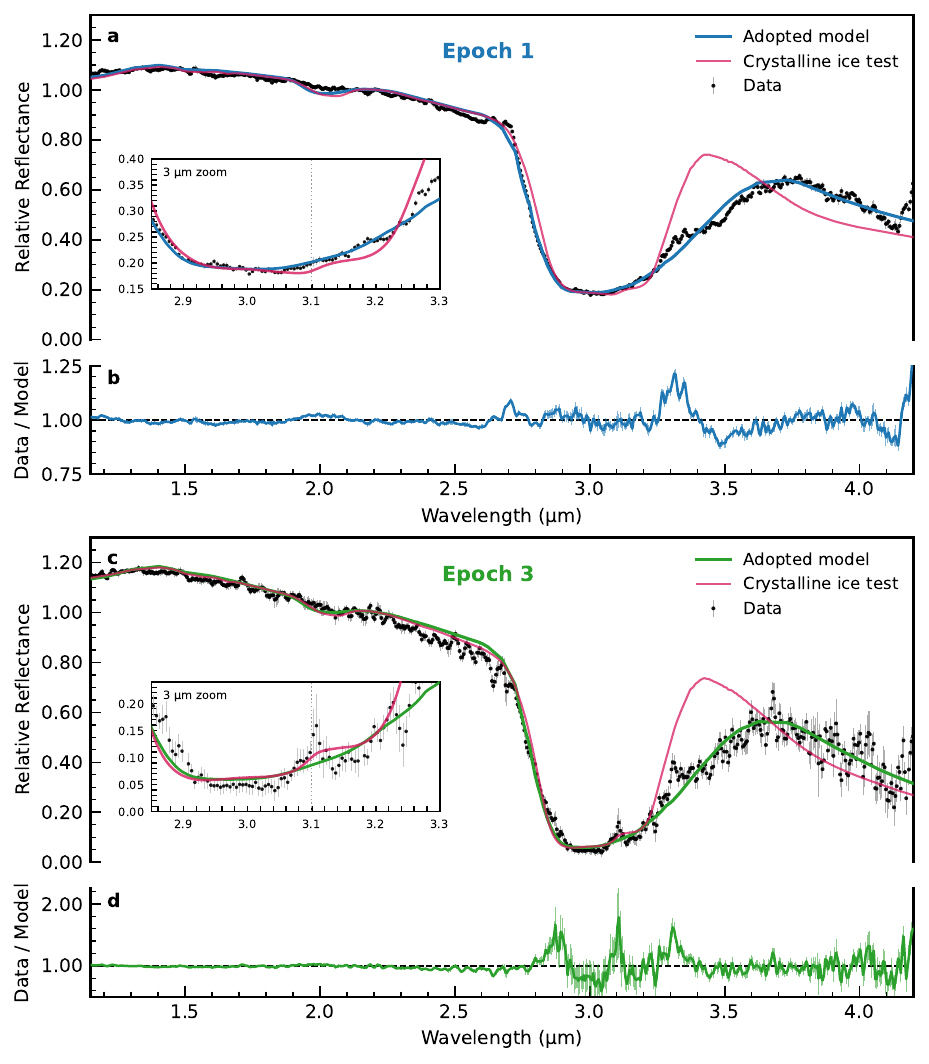}
\caption{Spectral modeling of the JWST/NIRSpec observations of 3I for Epochs~1 (a,b) and 3 (c,d). (a,c) The observed spectra (black) are compared with the preferred Warren-based best-fit models (blue for Epoch~1 and green for Epoch~3) and models computed using the crystalline-water-ice optical constants of \citet{Mastrapa2008,Mastrapa2009} (red). (b,d) Corresponding residuals. The Epoch~1 spectrum is best reproduced by fine-grained refractory-rich water-ice--pyroxene aggregates that suppress the weak 1.5 and 2.0~\textmu{}m absorptions while preserving the broad 3~\textmu{}m band. In contrast, the Epoch~3 spectrum favors an additional population of larger, less refractory-diluted water-ice--amorphous-carbon aggregates to reproduce the overall spectral morphology (continuum, weak overtone absorptions, and shape of the 3~\textmu{}m absorption band), including a Fresnel-like feature near 3.1~\textmu{}m. The inset panels emphasize the absence of a Fresnel-like structure in Epoch~1 and its presence in Epoch~3. Residual structure within the 2.6--2.8 and 3.2--3.6~\textmu{}m wavelength ranges is primarily associated with gaseous H$_2$O and organic emission features that were downweighted during the optimization.
\label{fig:best_fit}}
\end{figure*}
\begin{deluxetable*}{c c c c c c c c c c c c c c c c c}
\tablewidth{0pt}
\tablecaption{Mie-scattering models for Epoch~1 (obs001) and Epoch~3 (obs003).\label{tab:epoch_comparison}}
\setlength{\tabcolsep}{4pt}
\tablehead{
\colhead{Epoch} & \colhead{Model} &
\multicolumn{6}{c}{Type-A Aggregate} & \colhead{} &
\multicolumn{5}{c}{Type-B Aggregate} & \colhead{} & \colhead{} \\
\cline{3-8} \cline{10-15}
\colhead{} & \colhead{} &
\colhead{Comp.} & \colhead{$d_{\min}$} & \colhead{$d_{\max}$} & \colhead{$F$} & \colhead{$|\alpha|$} & \colhead{$\phi_{\rm base}$} &
\colhead{} &
\colhead{Comp.} & \colhead{$d_{\min}$} & \colhead{$d_{\max}$} & \colhead{$F$} & \colhead{$|\alpha|$} & \colhead{$\phi_{\rm base}$} &
\colhead{$\chi^2_\nu$} & \colhead{BIC} \\
\colhead{} & \colhead{} &
\colhead{} & \colhead{(\textmu{}m)} & \colhead{(\textmu{}m)} & \colhead{} & \colhead{} & \colhead{} &
\colhead{} &
\colhead{} & \colhead{(\textmu{}m)} & \colhead{(\textmu{}m)} & \colhead{} & \colhead{} & \colhead{} &
\colhead{} & \colhead{}
}
\startdata
1 & 1a & H$_2$O+px & 0.54 & 50$^{*}$ & 0.53 & 15.6 & 0.41
  & & px+AC & 4.14 & 50$^{*}$ & 0.47 & 45.5 & 0.82
  & 2.41 & -2027.6 \\
3 & 1a & H$_2$O+px & 0.41 & 50$^{*}$ & 0.72 & 12.8 & 0.28
  & & H$_2$O+AC & 1.54 & 50$^{*}$ & 0.28 & 47.1 & 0.95$^{*}$
  & 1.66 & -1223.4 \\
3 & 1b & H$_2$O+px & 0.44 & 50$^{*}$ & 0.59 & 50.0 & 0.32
  & & H$_2$O+AC & 1.53 & 50$^{*}$ & 0.41 & 50.0 & 0.95$^{*}$
  & -- & -- \\
3 & 2a & H$_2$O+px & 0.46 & 50$^{*}$ & 0.88 & 50.0 & 0.33
  & & px+AC & 4.81 & 50$^{*}$ & 0.12 & 50.0 & 0.93
  & 1.96 & -1038.7 \\
3 & 2b & H$_2$O+px & 0.45 & 50$^{*}$ & 0.87 & 50.0 & 0.40
  & & px+AC & 4.83 & 50$^{*}$ & 0.13 & 50.0 & 0.93
  & -- & -- \\
\enddata
\tablecomments{
For each particle population we report the minimum and maximum particle diameters, $d_{\min}=2r_{\min}$ and $d_{\max}=2r_{\max}$, the areal fraction, $F$, the differential particle-size distribution exponent, $\alpha$, and, for aggregate mixtures, the volume fraction of the first-listed component within the aggregate, $\phi_{\rm base}$. Model~1 denotes the preferred aggregate configuration for the corresponding epoch, whereas Model~2 applies the Epoch~1 aggregate configuration to the Epoch~3 spectrum. Models labeled ``a'' use the water-ice optical constants of \citet{Warren2008}, whereas models labeled ``b'' use the crystalline-water-ice optical constants of \citet{Mastrapa2008,Mastrapa2009}. Parameters marked with $^{*}$ were held fixed during the optimization. Reduced chi-square ($\chi^2_\nu$) and BIC values are reported only for models optimized over the full 1.15--4.2~\textmu{}m wavelength interval. Dashes indicate diagnostic models optimized over a restricted wavelength range.
}
\end{deluxetable*}
\subsection{Epoch 1: Refractory-Rich Water-Ice-Bearing Aggregates}

The best-fit model for Epoch~1 (Fig.~\ref{fig:best_fit}, panels~a and b, blue line; Table~\ref{tab:epoch_comparison}, Model~1a) requires Aggregate~A to consist of submicron- to micron-sized aggregates composed of water ice \citep{Warren2008} and amorphous pyroxene \citep[Mg/Fe ratio 50:50,][]{Dorschner1995}. The combined effects of small particle sizes and refractory mixing within the aggregates reduce the optical path length through the ice, thereby suppressing the weak 1.5 and 2.0-\textmu{}m water-ice bands while retaining a strong fundamental O--H stretching absorption near 3~\textmu{}m. Aggregate~A also contributes to the blue continuum across the 1.15--2.5~\textmu{}m wavelength range, with the continuum slope primarily controlled by particle size and the refractory volume fraction within the aggregates (Appendix~\ref{sec:appendix_modeling}, Figs.~\ref{fig:synthetic_spectra_purity} and \ref{fig:synthetic_spectra_grain_size}). Aggregate~B is composed of amorphous pyroxene and amorphous carbon \citep{Edoh1983} and primarily contributes to the overall continuum level. The resulting model reproduces both the near-infrared continuum and the broad 3~\textmu{}m absorption band. A small residual near 2.0~\textmu{}m indicates that the model slightly overestimates the weak overtone absorption, suggesting that the inferred refractory-rich water-ice-bearing aggregate component should be interpreted as an effective spectral representation of the coma grain population rather than a unique physical solution (Appendix~\ref{sec:appendix_modeling}). 

The robustness of this solution was tested through multiple optimization runs spanning a broad range of initial conditions. These consistently converged toward refractory-rich, submicron- to micron-sized water-ice-bearing aggregate solutions similar to the preferred model. Control models employing pure water ice or water ice mixed with amorphous carbon or olivine were also explored. These alternatives either overpredicted the 1.5 and 2.0~\textmu{}m absorptions or failed to reproduce the observed 3~\textmu{}m band, yielding larger Bayesian Information Criterion (BIC) values.

To investigate the sensitivity of the inferred solution to the adopted water-ice optical constants and the assumed ice phase and temperature, the preferred Epoch~1 solution was recomputed using the low-temperature crystalline water-ice optical constants of \citet{Mastrapa2008,Mastrapa2009} ($T=100$~K), while preserving the same aggregate configuration and model parameters derived for the Warren-based solution (Fig.~\ref{fig:best_fit}, panel~a, red line). The low-temperature crystalline optical constants of
\citet{Mastrapa2008,Mastrapa2009} intrinsically produce a stronger and more structured Fresnel feature than the optical constants of
\citet{Warren2008}, despite both representing crystalline water ice
(Appendix~{\ref{sec:appendix_fresnel}}, Fig.~\ref{fig:optical_constants}). Nevertheless, when incorporated within the inferred grain-size and refractory-mixing framework of the preferred Epoch~1 solution, neither optical-constant dataset produces a distinct Fresnel peak near 3.1~\textmu{}m. The resulting spectra therefore demonstrate that the absence of a pronounced Fresnel feature in Epoch~1 does not uniquely distinguish between crystalline and more amorphous-like spectral behavior. 

Spectral models employing the optical constants of \citet{Warren2008} provided a significantly better match to the broad 3.0--3.5~\textmu{}m absorption profile (Fig.~\ref{fig:best_fit}, panel~a, blue line) than equivalent models computed using the optical constants of \citet{Mastrapa2008,Mastrapa2009}, with particularly strong discrepancies between 3.3 and 3.5~\textmu{}m (Fig.~\ref{fig:best_fit}, panel~a, red line). These differences are inherent to the adopted optical-constant datasets: the Warren optical constants display an intrinsically broader 3~\textmu{}m water-ice absorption band than either the amorphous or crystalline datasets of \citet{Mastrapa2009} (Appendix~\ref{sec:appendix_fresnel}, Fig.~\ref{fig:optical_constants}). The origin of this discrepancy remains uncertain, and it is unclear whether it reflects differences in the underlying laboratory measurements and derivation of the optical constants or whether the various datasets capture distinct physical conditions, temperatures, and microphysical states of water ice relevant to astrophysical environments.
\subsection{Epoch 3: Emergence of a Second Ice-Rich Aggregate Population} 

The preferred model for Epoch~3 (Fig.~\ref{fig:best_fit}, panels~c and d, green line; Table~\ref{tab:epoch_comparison}, Model~1a) differs from the Epoch~1 solution in requiring both aggregate populations to contain water ice. Aggregate~A consists of water ice and pyroxene, as in Epoch~1, whereas Aggregate~B is composed of water ice mixed with amorphous carbon.

The water-ice--pyroxene aggregates primarily reproduce the overall spectral continuum and morphology of the 3~\textmu{}m absorption band, whereas the water-ice--amorphous-carbon aggregates help suppress the weak 1.5 and 2.0~\textmu{}m absorptions. The water-ice volume fraction within Aggregate~B was fixed at 0.95 because increasing refractory content progressively distorts the overall morphology of the 3~\textmu{}m absorption band. Nevertheless, the intrinsic 3~\textmu{}m morphology of water-ice--amorphous-carbon aggregates differs substantially from that observed in 3I (Appendix~\ref{sec:appendix_modeling}, Figs.~\ref{fig:synthetic_spectra_purity} and \ref{fig:synthetic_spectra_grain_size}), indicating that this component alone cannot reproduce the observations. The combined contribution of both aggregate populations is therefore required to simultaneously reproduce the continuum, weak overtone bands, and detailed 3~\textmu{}m band morphology.

To test whether the Epoch~3 spectrum requires a change in aggregate configuration relative to Epoch~1, we modeled Epoch~3 using the aggregate architecture adopted for Epoch~1 (Table~\ref{tab:epoch_comparison}, Model~2a), while retaining the same Warren water-ice optical constants and fitting the same 1.15--4.2~\textmu{}m wavelength interval. Model~2a provides a substantially poorer fit than the preferred Epoch~3 model ($\Delta{\rm BIC}=185$ in favor of Model~1a), despite having one additional free parameter. Importantly, this comparison is independent of the detailed Fresnel morphology near 3.1~\textmu{}m, because both models employ the Warren water-ice optical constants, which do not yield a pronounced Fresnel feature in the modeled spectra. Thus, within the physical models explored here, the preference for a change in the effective aggregate population at Epoch~3 is driven by the broader spectral constraints rather than by the local 3.1~\textmu{}m feature.

To assess whether the preferred Epoch~3 aggregate configuration can reproduce the subtle feature near 3.1~\textmu{}m when crystalline-water-ice optical constants are adopted, the Warren-based best-fit solution was re-optimized using the low-temperature crystalline-water-ice optical constants of \citet[][$T=100$~K]{Mastrapa2008,Mastrapa2009}. In light of the discussion above regarding the behavior of the Mastrapa optical constants between 3.3--3.5~\textmu{}m, and to isolate the sensitivity of the 3.1~\textmu{}m feature to the adopted water-ice optical constants, the re-optimization was restricted to shorter wavelengths up to 3.3~\textmu{}m, and the resulting model was subsequently extrapolated toward longer wavelengths (Fig.~\ref{fig:best_fit}, panel~c, red line; Table~\ref{tab:epoch_comparison}, Model~1b). 
The resulting model exhibits a more pronounced Fresnel feature near 3.1~\textmu{}m, although differences remain in the detailed morphology of the feature relative to the observations (Fig.~\ref{fig:best_fit}c, inset panel). These results highlight a trade-off between reproducing the overall 3~\textmu{}m band profile and preserving a Fresnel feature, reinforcing the sensitivity of the inferred solution to the adopted water-ice optical constants.

We performed the same Mastrapa-based re-optimization using the Epoch~1 aggregate architecture (Table~\ref{tab:epoch_comparison}, Model~2b), adopting the same 1.15--3.3~\textmu{}m wavelength interval as for Model~1b. Over this common fitting interval, Model~1b is strongly favored over Model~2b ($\Delta{\rm BIC}=194$, see Appendix~\ref{sec:appendix_modeling}, Fig.~\ref{fig:epoch3_model_comparison}). To quantify the model performance specifically around the 3.1~\textmu{}m feature, we additionally compared the residuals over 3.05--3.20~\textmu{}m. For the 29 spectral points within this interval, the residuals normalized by the observational uncertainties yield $\chi^2_{\rm local}=36.7$ for Model~1b and $\chi^2_{\rm local}=39.2$ for Model~2b, corresponding to RMS normalized residuals of 1.13 and 1.16, respectively. Thus, Model~1b provides a modestly better representation of the local 3.1~\textmu{}m morphology, while the much stronger preference for the Epoch~3 aggregate configuration arises from the broader spectral constraints rather than from the local feature alone.
\subsection{Temporal Evolution of the Coma Ice Population} 
Within this framework, the spectral evolution from Epoch~1 to Epoch~3 is primarily driven by the emergence of a second population of larger, micron-sized, ice-rich aggregates within the coma. Regarding the phase of the ice, the observations remain consistent with two scenarios: either the ice was crystalline during both epochs, with the Fresnel-like structure in Epoch~3 emerging primarily from changes in grain size and refractory mixing, or the ice evolved from a more amorphous-like state in Epoch~1 toward a more crystalline state by Epoch~3. The combined effects of grain size, refractory mixing, and uncertainties in the adopted optical constants prevent a unique distinction between these possibilities.

For both Epochs~1 and 3, the residuals show localized structure in wavelength regions affected by weak gas emission, particularly H$_2$O and organic features. These residuals are not interpreted as deficiencies of the solid-grain model because the affected spectral intervals were retained with inflated uncertainties and therefore contributed only weakly to the optimization. The posterior distributions for the preferred Epoch~1 and Epoch~3 solutions are presented in Appendix~\ref{sec:appendix_modeling} (Figs.~\ref{fig:mcmc_epoch1} and \ref{fig:mcmc_epoch3}), illustrating the principal parameter covariances and the overall stability of the inferred solutions. The strongest covariance occurs between the minimum particle size and the particle-size distribution slope of the ice-bearing aggregate component, reflecting the partial degeneracy between these parameters in controlling the effective optical path length through the aggregates.

\begin{figure*}[ht!]
\plotone{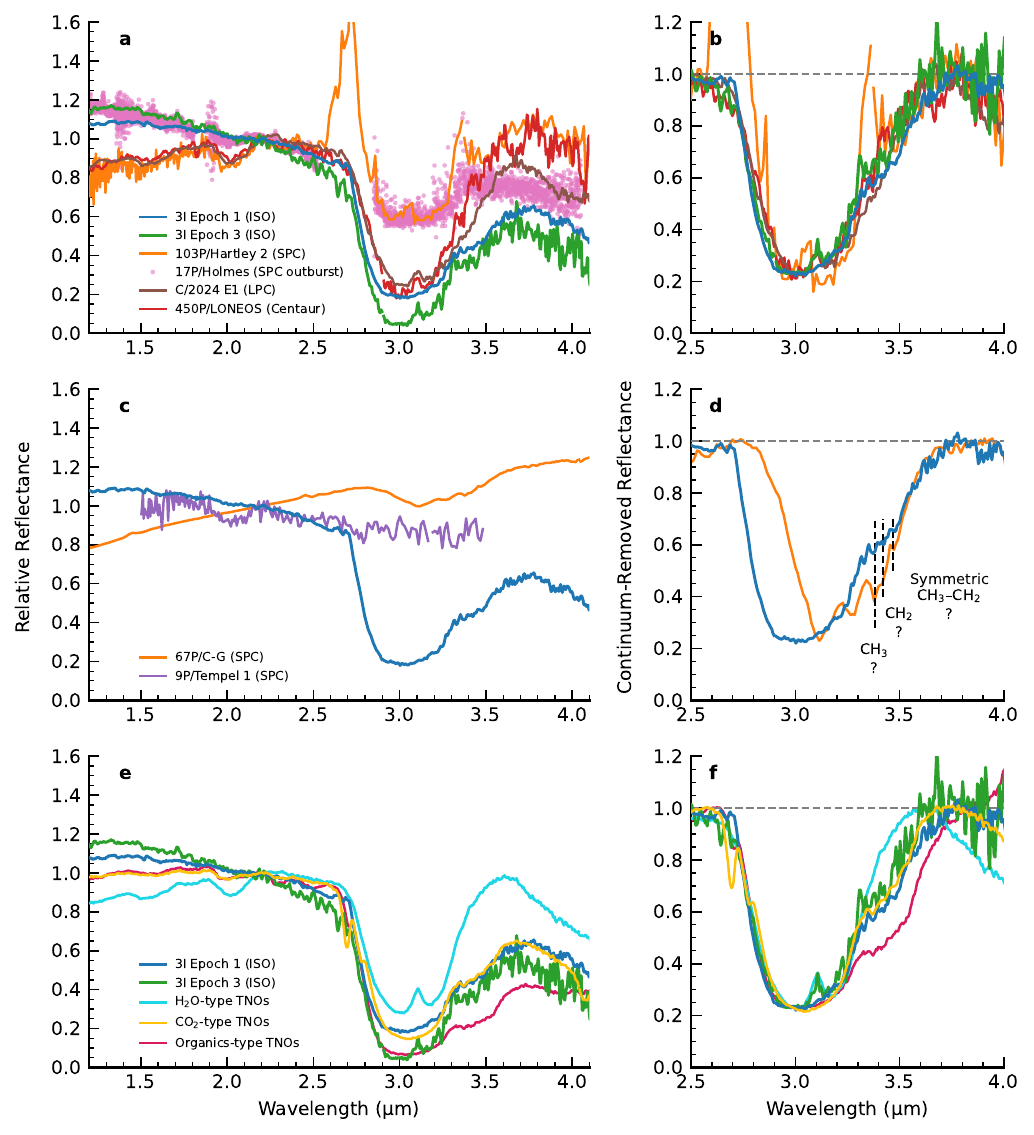}
\caption{Comparison of the water-ice absorption bands in 3I with those measured in cometary comae (a,b), cometary surfaces (c,d), and TNOs (e,f). Left panels show the spectra over the 1.2--4.1~\textmu{}m wavelength range, normalized between 2.18 and 2.22~\textmu{}m. Right panels show continuum-removed spectra scaled to the same 3~\textmu{}m band depth to facilitate comparison of the band morphology, following the approach of \citet{HarringtonPinto2023PSJ}. The continuum and band minimum were determined independently for each spectrum. For all spectra except 67P/Churyumov--Gerasimenko, the band minimum was measured between 2.96 and 3.00~\textmu{}m; for 67P, the minimum was measured between 3.11 and 3.13~\textmu{}m. (a,b) Near-infrared spectra of 3I (Epochs~1 and 3) compared with the water-ice-rich comae of 103P/Hartley~2 \citep{Protopapa2014}, C/2024~E1 \citep{Snodgrass2025}, 450P/LONEOS \citep{Schambeau2026}, and outbursting comet 17P/Holmes \citep{Yang2009}. While the overall morphology of the 3~\textmu{}m absorption band is similar among all objects, the 1.5 and 2.0~\textmu{}m overtone absorptions are substantially weaker in 3I. (c,d) Comparison with cometary surface spectra, including localized water-ice-rich exposures on 9P/Tempel~1 \citep[shown relative to a nearby non-ice region of the nucleus,][]{Sunshine2006Sci} and the average surface spectrum of 67P/Churyumov--Gerasimenko \citep{Raponi2020}. Dashed lines in panel~d indicate the positions of the aliphatic absorptions identified on 67P by \citet{Raponi2020}. (e,f) Comparison between 3I and representative spectra of the H$_2$O-type, CO$_2$-type, and organics-type TNO classes identified by the DiSCo survey \citep{Pinilla-Alonso2025,Holler2025}. The broad 3~\textmu{}m absorption in 3I resembles those observed across the three TNO classes. The Fresnel-like structure near 3.1~\textmu{}m in Epoch~3 is similar in both position and width to the Fresnel features observed in the water-ice-rich comae of C/2024~E1 and 450P/LONEOS \citep[panel b,][]{Snodgrass2025,Schambeau2026}, as well as in H$_2$O-rich TNOs \citep[panel f,][]{Pinilla-Alonso2025,Holler2025}.\label{fig:comparison}}
\end{figure*}
\section{Discussion} 
\subsection{Comparison to the properties of ice-bearing grains in Solar System comae}
Before comparing the inferred properties of water ice in 3I with those reported for Solar System active objects, we note an important distinction among the mixing treatments adopted in previous spectral studies. Water-ice-bearing coma spectra have commonly been modeled using an areal mixture of two spatially segregated particle populations: an ice-bearing population and refractory dust. In 103P/Hartley~2 (hereafter, Hartley~2), the ice-bearing population consists of pure water-ice grains \citep{Protopapa2014}. In C/2013~US10 (hereafter, US10), the ice-bearing particles themselves were modeled as aggregates containing a small refractory volume fraction (up to 1\%), with their optical properties calculated using effective-medium theory \citep{Protopapa2018}. The same effective-medium approach is adopted here for the ice-bearing aggregates in 3I, but with a substantially larger refractory contribution (Section~\ref{sec:modeling}). In the following, we therefore distinguish between pure water-ice grains (Hartley~2-like) and water-ice-bearing aggregates, with the latter described as refractory-poor for US10 and refractory-rich for 3I based on the refractory volume fraction within the aggregate ($\phi_{\rm ref}=1-\phi_{\rm H_2O}$; Section~\ref{sec:modeling}). 

450P/LONEOS (hereafter, 450P) represents a different modeling case: water ice and refractory material were combined through intimate mixing at the level of the single-scattering albedo rather than as separate populations in an areal mixture \citep{Schambeau2026}. Because inferred grain size and ice abundance are model-dependent, these properties should not be compared quantitatively on a one-to-one basis with those derived within the areal/effective-medium framework. Nevertheless, 450P provides an important point of comparison because it samples a different dynamical class and heliocentric-distance regime while exhibiting a water-ice spectral morphology remarkably similar to that observed in other ice-rich comae. A model of 450P computed within the same effective-medium framework adopted here would be required to place it quantitatively on the Hartley~2--US10--3I continuum in refractory content within the ice-bearing particles. 

The coma spectra of 3I differ markedly from previously reported detections of water ice in the comae of active objects, including the short-period comet Hartley~2 \citep[1~au;][]{Protopapa2014}, the long-period comet C/2024~E1 \citep[7~au;][]{Snodgrass2025}, the active Centaur 450P \citep[7~au;][]{Schambeau2026}, and the outbursting comet 17P/Holmes \citep[2.5~au;][]{Yang2009} (Fig.~\ref{fig:comparison}a). The morphology of the broad 3~\textmu{}m absorption band in 3I resembles that observed in these water-ice-rich comae, supporting its identification as the fundamental O--H stretching mode of H$_2$O ice (Fig.~\ref{fig:comparison}b). However, differences emerge in the weaker overtone and combination bands and in the spectral continua. Unlike Hartley~2, C/2024~E1, and 450P, which display readily detectable absorptions near 1.5 and 2.0~\textmu{}m in addition to the 3~\textmu{}m band, both Epoch~1 and Epoch~3 of 3I exhibit unusually weak or absent near-infrared overtone absorptions. The spectra of 3I also display substantially bluer near-infrared continua than the comparison comae (Fig.~\ref{fig:comparison}a). As demonstrated by the spectral modeling presented in Section~\ref{sec:modeling}, the combination of a blue continuum and suppressed overtone absorptions is most naturally explained by a substantial reduction in the effective optical path length through the ice, consistent with submicron- to micron-sized refractory-rich water-ice-bearing aggregates. Despite the difference in refractory content, the characteristic grain sizes inferred for 3I remain within the submicron-to-micron regime observed in Hartley~2 and other water-ice-bearing cometary comae \citep{Protopapa2014,Yang2014ApJ}. 

An intermediate case between 3I and more classical Hartley~2-like cometary comae may be represented by the outbursting comet 17P/Holmes, which shows a blue near-infrared continuum together with 2.0 and 3.0~\textmu{}m water-ice absorptions but no detectable 1.5~\textmu{}m band \citep{Yang2009}. Yang et al.\ interpreted these observations in terms of pure micron-sized water-ice grains physically separated from refractory dust, similar to the Hartley~2-like case, and identified the absence of the 1.5~\textmu{}m band as an unresolved problem. The refractory-rich water-ice-bearing aggregate interpretation inferred for 3I, potentially combined with particle-size distributions extending from submicron to micron scales, provides one possible mechanism capable of suppressing overtone absorptions, although dedicated modeling of the Holmes spectra would be required to assess whether a similar explanation applies in that case. 

Taken together, these comparisons indicate that fine-grained ice in the submicron-to-micron size regime is a recurrent property of water-ice-bearing comae of active objects across a broad range of heliocentric distances, dynamical classes, and activity states \citep{Sunshine2007Icar,Yang2009,Protopapa2014,Yang2014ApJ,Protopapa2018,Snodgrass2025,Schambeau2026}, while the ice-bearing particles span a range from pure water-ice grains to refractory-poor and refractory-rich aggregates. 3I therefore extends the diversity in refractory content observed within this otherwise recurrent fine-grained population. 

\subsection{Temporal evolution of 3I coma grains}
The evolution in the effective aggregate properties inferred between Epochs~1 and 3 may have two possible explanations. First, the substantially different thermal environments at 3.31~au and 5.65~au may affect the survival of ice-bearing particles in the coma, potentially favoring the detection of the water-ice--amorphous-carbon aggregate population at Epoch~3. Alternatively, the evolution may reflect material released from different depths within the nucleus, as progressive erosion through perihelion can expose increasingly deeper layers \citep{Frincke2026}. These scenarios cannot be distinguished with the present observations. Nevertheless, the ice-bearing grains inferred for 3I remain consistently within the submicron-to-micron size range across the observed epochs, while spanning a range of refractory contributions within the aggregates, consistent with the broader picture described above. 

\subsection{Comparison to the properties of cometary surfaces}
The spectrum of 3I also differs from previously characterized cometary surfaces. Although water ice on cometary nuclei spans a broad range of abundances, grain sizes, and mixing states, localized ice-rich exposures generally exhibit readily detectable near-infrared water-ice absorptions associated with substantially coarser ice than inferred for the coma of 3I, with characteristic grain sizes ranging from tens of microns to millimeters \citep{Sunshine2006Sci,Raponi2016,Filacchione2016Natur,Barucci2016}. The spectrum of water-ice-rich deposits on 9P/Tempel~1 shown in Fig.~\ref{fig:comparison}c (purple line) provides a representative example. By contrast, the average surface spectrum of 67P/Churyumov--Gerasimenko (67P) is dominated by refractory materials and displays additional absorptions near 3.38, 3.42, and 3.47~\textmu{}m attributed to aliphatic organics \citep[Fig.~\ref{fig:comparison}c, orange line;][]{Raponi2020}. 
The diversity in the surface composition of cometary nuclei and in the physical state of exposed surface ice has been interpreted as the result of near-surface evolution through thermal cycling, sublimation--recondensation, vapor transport, and sintering \citep{Filacchione2016Natur,Filacchione2019}. The underlying heterogeneity in bulk ice abundance, however, may itself be primordial, reflecting spatial variations in the concentration of ice-bearing material within the nucleus rather than necessarily differences in the physical form of the ice itself \citep{Ciarniello2022}. 

An important exception to the generally coarser ice observed in localized surface exposures is the 1--3~\textmu{}m frost detected by \textit{Rosetta} at Hapi on 67P \citep{DeSanctis2015Nature}. This fine-grained surface population has been interpreted as a secondary product of H$_2$O sublimation, vapor transport, and recondensation within the shallow subsurface. By contrast, the recurrence of similarly fine-grained ice in the comae of active objects at heliocentric distances extending to $\sim$7~au \citep{Schambeau2026,Snodgrass2025}, where H$_2$O sublimation is inefficient, cannot readily be explained by the same process. Thus, despite their similar characteristic grain sizes, the Hapi frost and the recurrent fine-grained coma populations likely reflect different physical origins. 

The Epoch~1 spectrum also exhibits weak residual structure across the 3.3--3.6~\textmu{}m region that is not fully reproduced by the preferred water-ice model and may indicate contributions from aliphatic organic absorptions similar to those identified on the surface of 67P \citep{Raponi2020}; however, the present data do not permit a unique compositional interpretation (Fig.~\ref{fig:comparison}d and Appendix~\ref{sec:appendix_organics}). 

\subsection{Comparison to the properties of TNO surfaces}
The comparison with TNO surface spectra is particularly instructive (Fig.~\ref{fig:comparison}e,f). After continuum removal, the 3~\textmu{}m absorption band of 3I closely resembles those of the H$_2$O-type, CO$_2$-type, and organics-type TNO classes identified by the DiSCo survey \citep{Pinilla-Alonso2025,Holler2025}. The largest differences among the DiSCo classes in the 2.8--3.8~\textmu{}m wavelength range occur longward of $\sim$3.3~\textmu{}m, where the CO$_2$-type and organics-type spectra exhibit additional absorption structure attributed to organic species. In particular, the continuum-removed spectra of both Epochs~1 and 3 display an overall band morphology similar to that of the CO$_2$-type class over 2.8--3.8~\textmu{}m. This contrasts with previous suggestions of an analogy between 3I and the organics-rich ``cliff-type'' TNO spectra \citep{Lisse_2026ApJ_3I}. The Fresnel-like structure near 3.1~\textmu{}m in Epoch~3 is similar in both position and width to the Fresnel structure observed in H$_2$O-type TNOs (Fig.~\ref{fig:comparison}f). 

In the DiSCo interpretation, the 3~\textmu{}m absorption observed in H$_2$O-type TNOs is attributed primarily to water ice, supported by the simultaneous presence of the 1.5 and 2.0~\textmu{}m absorptions and the characteristic Fresnel reflection feature near 3.1~\textmu{}m. In contrast, the 3~\textmu{}m absorptions observed in the CO$_2$-type and organics-type classes are interpreted as arising from O--H-bearing materials such as organics and methanol, with a possible minor contribution from amorphous water \citep{Pinilla-Alonso2025}, given the absence of obvious water-ice features near 1.5 and 2.0~\textmu{}m. However, the spectra and modeling of 3I demonstrate that a similar 3~\textmu{}m band morphology can be reproduced by water-ice-bearing aggregates in which the 1.5 and 2.0~\textmu{}m overtone bands are strongly suppressed by small grains and refractory mixing. The presence of a Fresnel-like structure near 3.1~\textmu{}m in Epoch~3 is also consistent with this water-ice interpretation. These results therefore raise the possibility that water ice within refractory-rich aggregates contributes to the 3~\textmu{}m absorption morphology of the CO$_2$-type and organics-type TNO classes, with the long-wavelength wing of the water-ice band further modified by superposed absorptions from organic species. 

\subsection{Synthesis}
The similarity between the 3~\textmu{}m band morphology of 3I and the CO$_2$-type and organics-type TNO classes, together with the suppression of the 1.5 and 2.0~\textmu{}m overtone bands and the blue near-infrared continua observed in several members of these classes, suggests that submicron- to micron-sized water-ice-bearing aggregates containing refractory material is a pervasive component of icy planetesimals. When considered together with the relatively pure, micron-sized water ice mechanically excavated from the subsurface of 9P/Tempel~1 \citep{Sunshine2006Sci} and previously characterized ice-rich comae, which are generally interpreted as sampling relatively unprocessed water ice from the interiors of Solar System planetesimals \citep[e.g.,][]{Protopapa2014,Protopapa2018,Sunshine2021}, these results suggest that the recurrent submicron-to-micron grain-scale character of water ice may reflect a property of the interiors of icy planetesimals formed in different protoplanetary disks, rather than being primarily a product of near-surface processing. The effective particle sizes inferred from coma spectra, however, should not be interpreted as a direct measurement of the pristine grain-size distribution within the nucleus, since the observed particles may represent individual grains, fragments, or subunits of larger aggregates. Within this limitation, water ice in planetesimals appears to span a broader continuum of physical states than previously recognized, from relatively pure ice grains to water-ice-bearing aggregates with varying refractory contributions. This interpretation is consistent with recent observations and modeling of protoplanetary and debris disks, which reveal micron-sized icy grains, compositionally complex ice-bearing populations, and a strong dependence of near-infrared water-ice spectral morphology on grain-scale properties \citep{Tazaki2021,Sturm2023,Xie2025}, supporting the idea that such grain structures may be inherited from the planetesimal-building environments in which these bodies formed. 

We emphasize, however, that comparisons with TNO surface spectra should be treated with caution, since the observed spectral properties of TNOs reflect both their primordial heritage and subsequent evolutionary processing of their surfaces \citep{Brunetto2025ApJL,Wong2025}. While some spectral characteristics may preserve information about the physical state of primordial icy materials, others likely reflect irradiation, impact excavation, and thermal processing, all of which can modify the abundance, distribution, and spectral expression of volatile species such as CO \citep{Henault2025} and CO$_2$ \citep{Protopapa2024}. We interpret the similarities between 3I, water-ice-bearing comae, and the DiSCo TNO classes as evidence that these populations may preserve aspects of the physical state of water ice established during planetesimal formation despite subsequent evolutionary processing. The spectral similarities discussed here therefore do not imply identical parent-disk compositions. Indeed, other compositional differences reported for 3I suggest that its natal disk may have differed chemically from the Solar Nebula \citep{Cordiner2026}. Nevertheless, the comparison of water-ice properties across these populations suggests that submicron-to-micron grain-scale properties, spanning a range of refractory contributions, may be a common characteristic of icy planetesimals formed in distinct protoplanetary environments.

\section{Summary and Outlook} 

JWST/NIRSpec observations of 3I reveal water-ice-bearing coma grains with spectral properties distinct from those previously characterized in water-ice-rich Solar System comae. The spectra exhibit a strong 3~\textmu{}m H$_2$O-ice absorption band but unusually weak 1.5 and 2.0~\textmu{}m bands, together with blue near-infrared continua and, in Epoch~3, the presence of a Fresnel-like structure near 3.1~\textmu{}m. Our modeling shows that these properties can be reproduced by submicron- to micron-sized water-ice-bearing aggregates containing refractory material. Such mixtures reduce the effective optical path length through the ice, suppressing the overtone and combination bands at 1.5 and 2.0~\textmu{}m while preserving the fundamental O--H stretching absorption near 3~\textmu{}m. The Fresnel-like structure near 3.1~\textmu{}m in Epoch~3 is also consistent with the Fresnel structure observed in H$_2$O-type TNOs \citep{Pinilla-Alonso2025} and the ice-rich comae of C/2024~E1 and 450P/LONEOS \citep{Snodgrass2025,Schambeau2026}. At the same time, the morphology of the broad 3~\textmu{}m absorption remains largely unchanged between epochs, supporting a common water-ice origin for the band both before and after perihelion. Independent evidence for an extended source of water vapor in the coma, consistent with sublimating icy grains contributing to the observed water production \citep{Roth2026,Cordiner2025,Cordiner2026,Bockelee-Morvan2026,Belyakov2026ApJ}, further supports the interpretation of the spectral morphology in terms of water ice. 

Through the comparison of an interstellar object with Solar System ice-rich comae and TNOs, we may be identifying common physical properties of the icy building blocks of planetesimals across planetary systems. In this picture, water ice spans a continuum of physical states ranging from pure water-ice grains to water-ice-bearing aggregates with varying refractory contributions across a recurrent submicron-to-micron grain-scale range. Expanding the spectroscopic characterization of cometary coma ice across different dynamical classes and heliocentric distances, together with future observations of interstellar objects from diverse natal environments, offers the opportunity to constrain the physical state of water ice that may be inherited during planetesimal formation. 

The sensitivity of the inferred grain properties to the adopted water-ice optical constants underscores the need for renewed laboratory investigations of water ice spanning a broad range of temperatures, phases, and microphysical states. As evidence continues to accumulate that some spectral properties of water ice observed in protoplanetary disks, small bodies, and interstellar objects may preserve information about the physical state of ice established during planetesimal formation, improved laboratory optical constants will be essential for robustly linking observed spectral morphologies to the physical state of the ice across these diverse environments. 

\begin{acknowledgments}
This work is based on observations made with the NASA/ESA/CSA James Webb Space Telescope. The data were obtained from the Mikulski Archive for Space Telescopes at the Space Telescope Science Institute, which is operated by the Association of Universities for Research in Astronomy, Inc., under NASA contract NAS 5-03127 for JWST. These observations are associated with program \#5094. Support for US investigators in program \#5094 was provided by NASA through a grant from the Space Telescope Science Institute. The dataset analyzed in this work is available through MAST at \dataset[DOI: 10.17909/krfj-zw46]{https://doi.org/10.17909/krfj-zw46}.

SBC, MAC, SNM and NXR were supported by the NASA Planetary Science Division Internal Scientist Funding Program through the Fundamental Laboratory Research work package (FLaRe).

MND's work is funded by the European Union. Views and opinions expressed are however those of the author(s) only and do not necessarily reflect those of the European Union or the European Research Council Executive Agency. Neither the European Union nor the granting authority can be held responsible for them. This work is supported by ERC grant PSII (DOI: 10.3030/101230593).

DF conducted this research at the Jet Propulsion Laboratory, California Institute of Technology, under a contract with the National Aeronautics and Space Administration (NASA) (80NM0018D0004).

DZS acknowledges funding support from JWST GO 5959, which was provided by NASA through a grant from the Space Telescope Science Institute.

S.P. thanks Bin Yang for sharing the spectrum of 17P/Holmes, Jessica M. Sunshine for sharing the spectrum of 9P/Tempel~1, and Bryan J. Holler for providing the average spectra of the TNO spectral classes used in Fig.~\ref{fig:comparison}.

We thank the anonymous referee for their thoughtful and constructive comments, which helped improve the manuscript.
\end{acknowledgments}





%
\facilities{JWST}

\software{NumPy \citep{Harris2020NumPy}, 
Matplotlib \citep{Hunter2007Matplotlib},
SciPy \citep{Virtanen2020NatSciPy},
\textcolor{black}{Astropy \citep{astropy:2013, astropy:2018, astropy:2022}},
Photutils \citep{Bradley_2024_photutils}, 
emcee \citep{Foreman-Mackey2013emcee}, 
SBPy \citep{Mommert2019}.}


\appendix
\section{Water Ice Fresnel Reflection Feature}\label{sec:appendix_fresnel}
\begin{figure*}[ht!]
\plotone{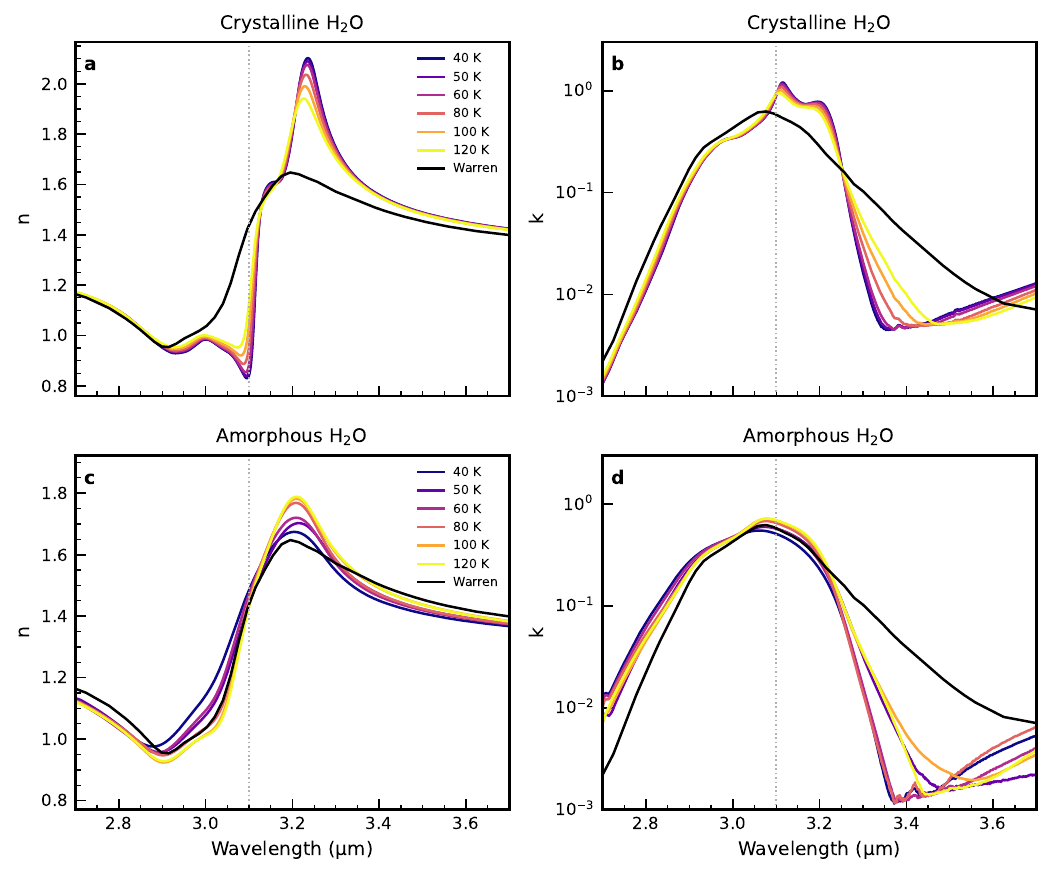}
\caption{
Comparison of the real ($n$; left panels) and imaginary ($k$; right panels) parts of the complex refractive index of crystalline (top) and amorphous (bottom) water ice near the 3~\textmu{}m O--H stretching band. Colored curves correspond to the temperature-dependent optical constants of \citet{Mastrapa2009} spanning 40--120~K, while the black curves show the \citet{Warren2008} compilation, nominally characteristic of crystalline water ice near 266~K. The vertical dotted line marks 3.1~\textmu{}m. The low-temperature crystalline optical constants exhibit strong anomalous dispersion in $n$ near 3.1~\textmu{}m together with resolved substructure in $k$, both of which are substantially weaker and smoother in amorphous ice. The warmer, lower-resolution \citet{Warren2008} compilation displays smoother behavior in $n$ and $k$ than the low-temperature crystalline optical constants of \citet{Mastrapa2009} and a broader $k$ profile across the 3~\textmu{}m absorption complex than either the crystalline or amorphous \citet{Mastrapa2009} datasets. These differences in $n$ and $k$ translate into reflectance spectra in which the \citet{Warren2008} optical constants produce a broader 3~\textmu{}m absorption profile and a weaker Fresnel reflection feature than the low-temperature crystalline optical constants of \citet{Mastrapa2009}, with important implications for interpreting the observed 3~\textmu{}m spectral morphology of 3I.
\label{fig:optical_constants}}
\end{figure*}

A Fresnel reflection feature near 3.1~\textmu{}m arises from the spectral
behavior of the real and imaginary parts of the complex refractive index,
$n$ and $k$, across the strong fundamental O--H stretching absorption of
water ice. The morphology of this feature depends on the phase of the ice:
amorphous ice presents a broader and smoother feature, whereas crystalline
water ice displays a more pronounced and structured reflection, driven by
a larger excursion in $n$ and a more complex profile in $k$ near
3.1~\textmu{}m \citep{Hansen2004,Mastrapa2009}. Figure~\ref{fig:optical_constants}
compares the real and imaginary parts of the complex refractive index near
the 3~\textmu{}m absorption complex and illustrates the physical origin of
the Fresnel reflection feature near 3.1~\textmu{}m, together with the
dependence of its morphology and that of the broader 3~\textmu{}m band
profile on ice phase, temperature, and the adopted optical constants.
\subsection{Robustness of the Epoch~3 Spectral Structure near 3.1~\textmu{}m}
\label{sec:appendix_fresnel_robustness}

\begin{figure*}[ht!]
\includegraphics[width=\textwidth]{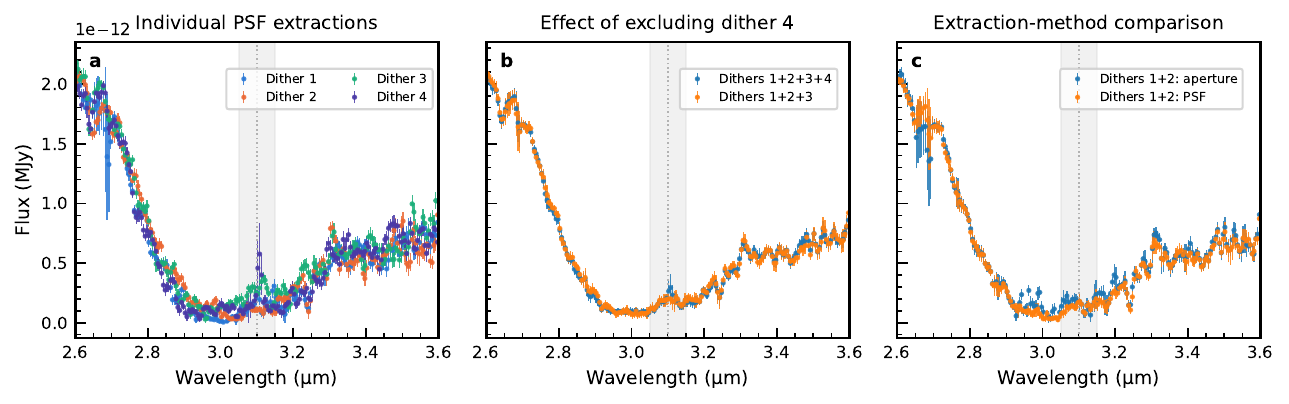}
\caption{Robustness tests for the spectral structure near 3.1~\textmu{}m
in the Epoch~3 spectrum. (a) Individual PSF-extracted spectra for the four
dithers. (b) Nominal four-dither mean spectrum compared with the mean
constructed from dithers 1--3. (c) PSF and aperture extractions constructed
independently from dithers 1 and 2. Error bars show the corresponding
spectral uncertainties. The vertical dotted line marks 3.1~\textmu{}m and
the shaded region highlights the wavelength interval surrounding the
spectral structure discussed in the text.}
\label{fig:fresnel_robustness}
\end{figure*}

Because the flux approaches zero near the center of the
3~\textmu{}m absorption band, we tested the robustness of the subtle
structure near 3.1~\textmu{}m in Epoch~3 against individual dithers and the
adopted spectral extraction methodology. Figure~\ref{fig:fresnel_robustness}
compares the four individual PSF-extracted spectra, the nominal four-dither
mean spectrum with a mean constructed from dithers 1--3 only, and PSF and
aperture extractions constructed independently from dithers 1 and 2.
Excluding dither 4 preserves the broader spectral morphology near
3.1~\textmu{}m, although dither 4 enhances the sharper local maximum.
The independent aperture extraction is noisier near the absorption minimum
but exhibits broadly consistent morphology. These tests indicate that the
broader structure near 3.1~\textmu{}m is not dependent on a single dither
nor on the PSF-based extraction, while the detailed shape of the local
feature is less robust.
\section{Spectral Modeling}\label{sec:appendix_modeling}
\subsection{Coma Reflectance Model}
The coma reflectance model summarized in Section~\ref{sec:modeling}
is based on an areal mixture of two particle populations.
For each population, the equivalent single-scattering albedo,
$w$, was computed by integrating the scattering and extinction
efficiencies, $Q_{\rm sca}$ and $Q_{\rm ext}$, over a differential
particle-size distribution, $n(r)$, between the minimum and maximum particle radii, $r_{\min}$ and $r_{\max}$:

\begin{equation}
w(\lambda)=
\frac{
\int_{r_{\min}}^{r_{\max}}
n(r)\:\pi r^2 \:Q_{\mathrm{sca}}(r,\lambda)\:dr
}{
\int_{r_{\min}}^{r_{\max}}
n(r)\:\pi r^2\:Q_{\mathrm{ext}}(r,\lambda)\:dr
}.
\end{equation}

Here, $n(r)\,dr$ represents the number of particles with radii between $r$ and $r+dr$, and $n(r)$ was assumed to follow a power law of the form $n(r)\propto r^{-\alpha}$. The scattering and extinction efficiencies were computed using Mie theory \citep{BohrenHuffman1983} from the wavelength-dependent complex refractive index, $m(\lambda)=n(\lambda)+ik(\lambda)$. For aggregate grains, the complex refractive index was derived using an effective-medium approximation through the Bruggeman mixing formula \citep{BohrenHuffman1983}, thereby representing sub-particle-scale mixing among the constituent materials. In this formalism, the constituent materials contribute collectively to the optical response of the aggregate while retaining their individual material properties.

The observed reflectance was modeled as

\begin{equation}
R(\lambda)=F_{A}R_{A}+(1-F_{A})R_{B},
\end{equation}
where A and B represent the two grain populations within the coma, $F_{A}$ is the areal fraction of population~A, and $R_{A}$ and $R_{B}$ are the reflectances of the individual populations, approximated using the semi-infinite diffuse-reflectance formulation \citep{Hapke2012}:

\begin{equation}
R_{i}(\lambda)=\frac{1-\sqrt{1-w_{i}(\lambda)}}{1+\sqrt{1-w_{i}(\lambda)}}.
\end{equation}

Although the final reflectance is expressed using the semi-infinite diffuse approximation of \citet{Hapke2012}, the present implementation differs from traditional surface-oriented Hapke slab models in that the single-scattering albedo is derived from Mie-computed scattering properties integrated over a coma particle-size distribution. The resulting formulation approximates the scattering behavior of an optically thin coma composed of discrete aggregate grains rather than that of a semi-infinite particulate surface.
\subsection{Optimization procedure}
We first performed a Levenberg–Marquardt (LM) minimization to identify an initial solution in parameter space. This solution was then used to initialize a Markov Chain Monte Carlo (MCMC) analysis aimed at exploring the posterior distributions and quantifying parameter uncertainties. Convergence of the MCMC chains was assessed through both autocorrelation-time analysis and visual inspection to verify that the chains were well mixed and stationary. After convergence was achieved, we performed a final LM optimization initialized from the median posterior values obtained from the MCMC analysis to refine the best-fit solution and minimize residual structure. Relative model comparison was performed using the Bayesian Information Criterion (BIC), which allows different aggregate configurations to be evaluated while accounting for differences in model complexity.
\subsection{Effects of Aggregate Composition and Optical Constants on Water-Ice Spectra}

\begin{figure*}[ht!]
\plotone{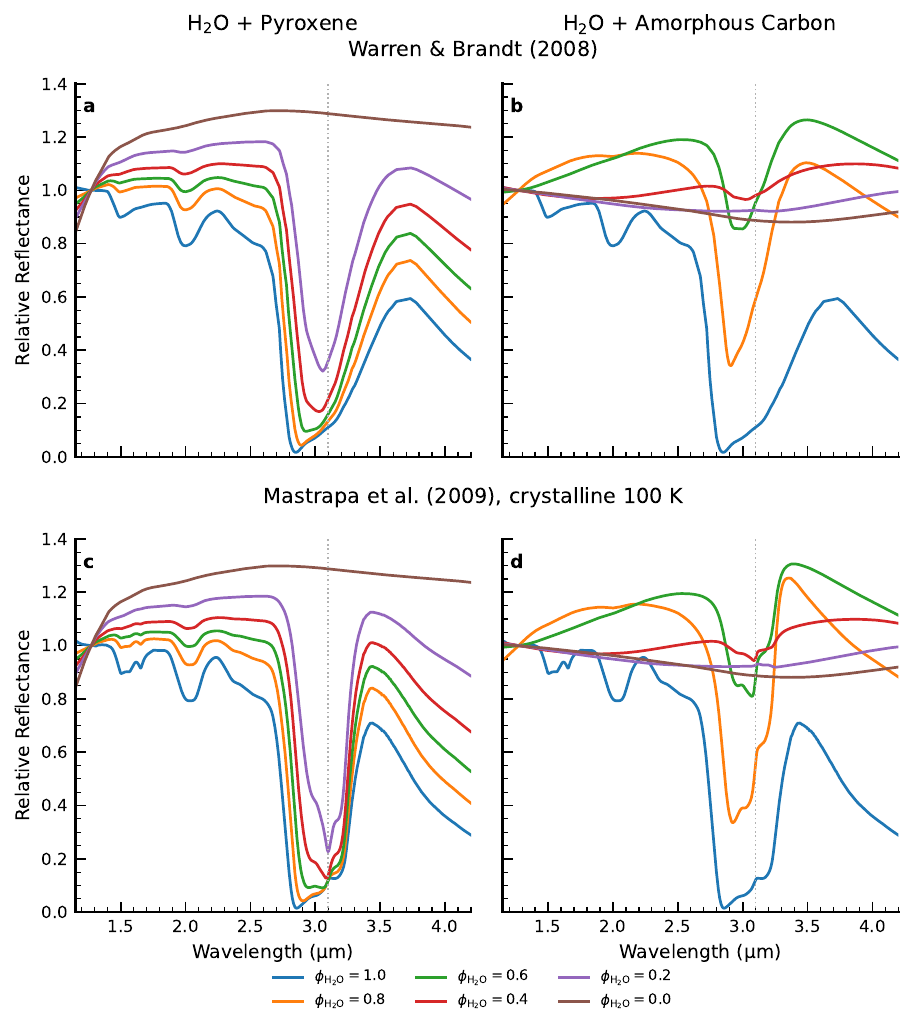}
\caption{
Synthetic reflectance spectra illustrating the effects of refractory composition on the near-infrared spectral morphology of water ice. Left panels show aggregate particles consisting of water ice and amorphous pyroxene, whereas right panels show aggregates of water ice and amorphous carbon. The top and bottom rows adopt the water-ice optical constants of \citet{Warren2008} and \citet{Mastrapa2008,Mastrapa2009}, respectively. In all cases, the spectra illustrate a continuum extending from pure water-ice grains ($\phi_{\mathrm{H_2O}}=1$) to refractory-rich aggregates ($\phi_{\mathrm{H_2O}}<1$). Refractory association, through both the abundance and composition of refractory material within the aggregate, suppresses the 1.5 and 2.0~\textmu{}m overtone and combination absorptions while simultaneously modifying the morphology of the 3~\textmu{}m absorption complex and the strength of the Fresnel reflection feature near 3.1~\textmu{}m. Weak or absent overtone bands therefore do not necessarily imply the absence of water ice, but may instead reflect reduced optical path lengths within water-ice-bearing aggregates containing refractory material.
\label{fig:synthetic_spectra_purity}}
\end{figure*}

\begin{figure*}[ht!]
\centering
\includegraphics[width=0.8\textwidth]{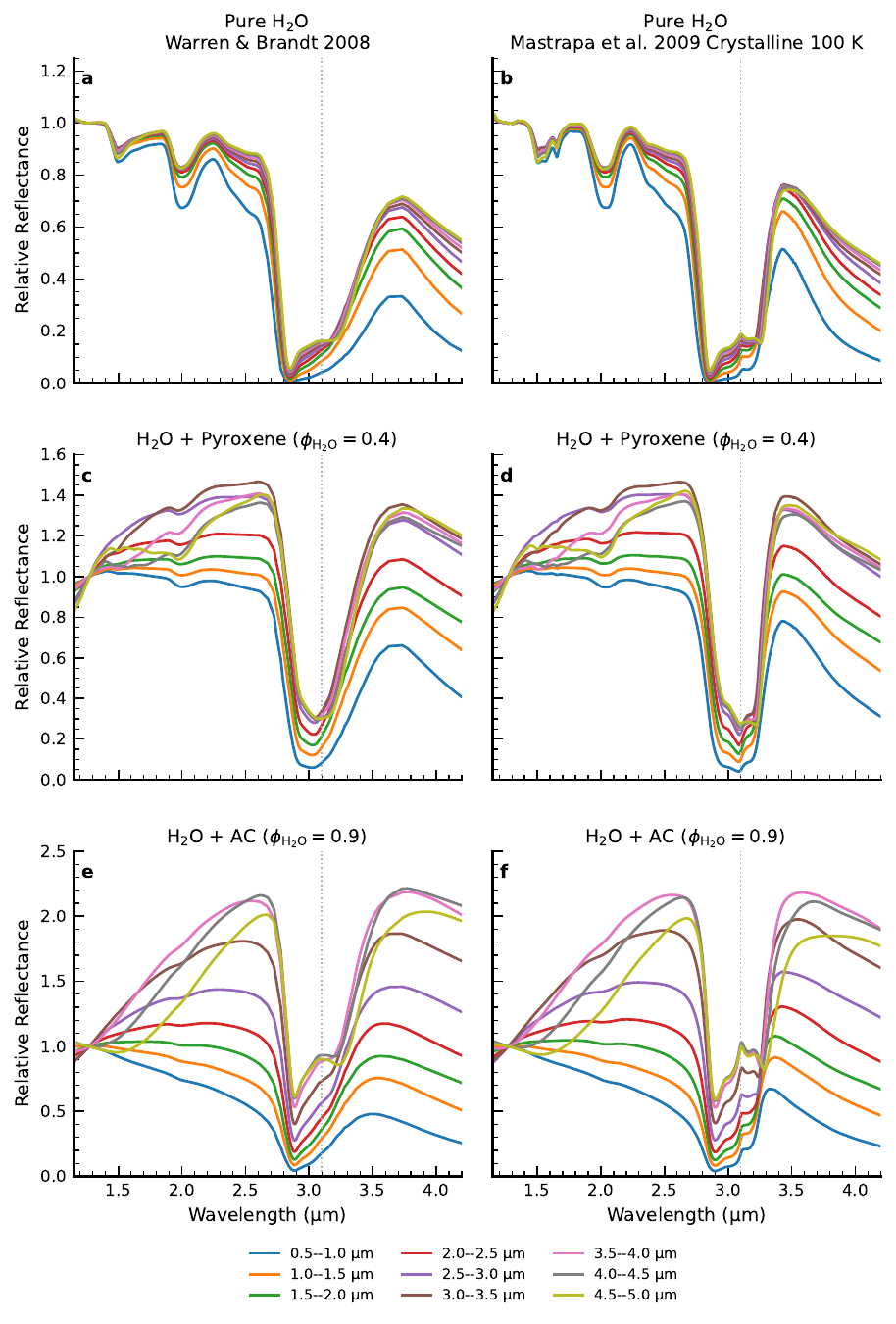}
\caption{
Synthetic reflectance spectra illustrating the influence of particle size, aggregate composition, and adopted water-ice optical constants on the near-infrared spectral morphology of water ice. Panels~a and b show pure water ice, panels~c and d show aggregates of water ice and amorphous pyroxene with $\phi_{\mathrm{H_2O}}=0.4$, and panels~e and f show aggregates of water ice and amorphous carbon with $\phi_{\mathrm{H_2O}}=0.9$. Left and right columns adopt the water-ice optical constants of \citet{Warren2008} and low-temperature crystalline water ice from \citet{Mastrapa2008,Mastrapa2009}, respectively. Variations in particle size modify the continuum slope, the relative strength of the water-ice absorption bands, and the detailed morphology of the 3~\textmu{}m absorption complex. The strength and visibility of the Fresnel reflection feature near 3.1~\textmu{}m depend strongly on both aggregate composition and particle size, demonstrating that its absence does not necessarily imply amorphous water ice.
\label{fig:synthetic_spectra_grain_size}}
\end{figure*}

To investigate the spectral behavior of water ice as a function of optical constants, aggregate composition, and grain size, we employed the spectral-modeling parameterization described in Section~\ref{sec:modeling} and computed synthetic reflectance spectra representative of the observed coma. Two sets of optical constants were considered for water ice: the water-ice optical constants of \citet{Warren2008}, nominally characteristic of crystalline ice near 266~K, and the low-temperature crystalline optical constants of \citet{Mastrapa2008,Mastrapa2009} at 100~K (Fig.~\ref{fig:optical_constants}). To explore the effects of aggregate composition, synthetic spectra were computed for aggregate particles consisting of water ice and either amorphous pyroxene \citep[Mg/Fe ratio 50:50][]{Dorschner1995} or amorphous carbon \citep{Edoh1983}, spanning a continuum from pure water-ice grains ($\phi_{\mathrm{H_2O}}=1$) to refractory-rich aggregates ($\phi_{\mathrm{H_2O}}<1$). Narrow particle-size distributions with $\alpha=1$ and aggregate sizes in the range 1.5--2~\textmu{}m were considered (Fig.~\ref{fig:synthetic_spectra_purity}). Finally, we explored the influence of particle size for pure water ice and for representative water-ice-bearing aggregates consisting of water ice and amorphous pyroxene with $\phi_{\mathrm{H_2O}}=0.4$, and water ice and amorphous carbon with $\phi_{\mathrm{H_2O}}=0.9$. Narrow particle-size distributions with a power-law exponent of $\alpha=1$ and aggregate grain sizes spanning 0.5--5~\textmu{}m were adopted, corresponding to the range most relevant for interpreting the 3I observations (Fig.~\ref{fig:synthetic_spectra_grain_size}). 

Pure water-ice grains display strong absorption bands at 1.5, 2.0, and 3.0~\textmu{}m simultaneously. Refractory mixing within aggregates substantially reduces the effective optical path length through the ice, with increasing refractory abundance leading to stronger suppression of the weaker 1.5 and 2.0~\textmu{}m overtone and combination bands while preserving the fundamental O--H stretching absorption near 3~\textmu{}m. The degree of suppression depends not only on the refractory volume fraction within the aggregate but also on the specific optical properties of the refractory component itself, with amorphous carbon producing substantially stronger attenuation of the overtone bands than pyroxene for comparable water-ice abundances (Fig.~\ref{fig:synthetic_spectra_purity}). In addition, both the detailed morphology of the 3~\textmu{}m absorption complex and the continuum slope across the full wavelength range depend sensitively on refractory composition and volume fraction within the aggregate and particle size (Figs.~\ref{fig:synthetic_spectra_purity} and \ref{fig:synthetic_spectra_grain_size}).

\begin{figure*}[ht!]
\centering
\includegraphics[width=0.8\textwidth]{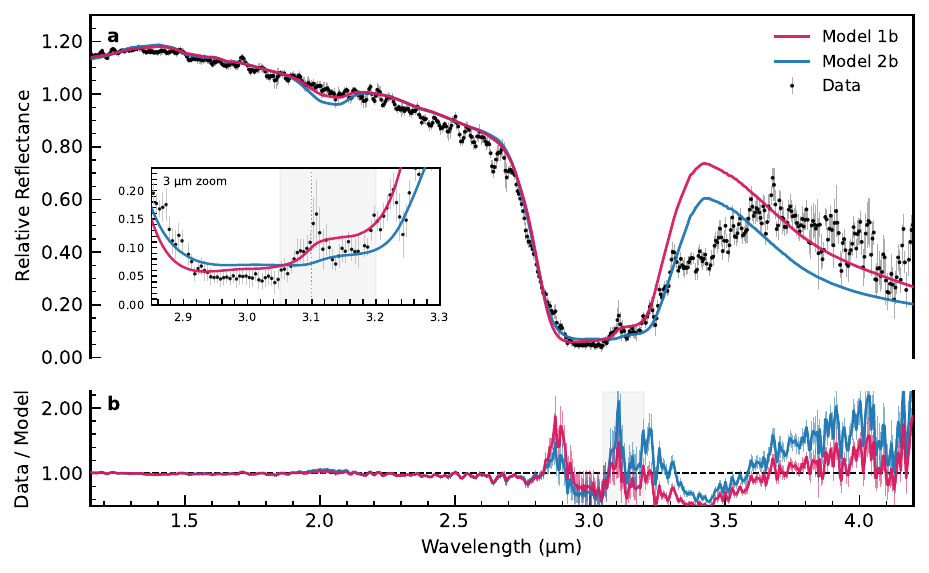}
\caption{Comparison of the Epoch~3 spectrum with the two Mastrapa-based aggregate configurations. Model~1b adopts the preferred Epoch~3 architecture, in which Aggregate~A consists of water ice and pyroxene and Aggregate~B consists of water ice and amorphous carbon, whereas Model~2b adopts the Epoch~1 architecture, in which Aggregate~B is ice-free. Both models use the low-temperature crystalline-water-ice optical constants of \citet{Mastrapa2008,Mastrapa2009} and were optimized over 1.15--3.3~\textmu{}m; the models are extrapolated to longer wavelengths for display. The upper panel shows the observed spectrum and model spectra, with the inset highlighting the 3~\textmu{}m absorption complex. The lower panel shows the corresponding data-to-model residuals, including the propagated observational uncertainties. The shaded region marks 3.05--3.20~\textmu{}m, the interval used for the local residual comparison (see main text for details), and the vertical dotted line marks 3.1~\textmu{}m.
\label{fig:epoch3_model_comparison}}
\end{figure*}

The morphology of the 3.1~\textmu{}m Fresnel reflection feature depends on the adopted water-ice optical constants. Models computed using the optical constants of \citet{Warren2008} yield a broader and weaker Fresnel reflection feature relative to the more structured profile obtained using the low-temperature crystalline-water-ice optical constants of \citet{Mastrapa2009}. The Fresnel structure remains visible in aggregates containing amorphous carbon and shows a strong dependence on grain size, whereas it becomes substantially weakened in pyroxene-rich aggregates (Figs.~\ref{fig:synthetic_spectra_purity} and \ref{fig:synthetic_spectra_grain_size}). The Epoch~3 comparison between the preferred aggregate architecture and the Epoch~1 aggregate architecture, both computed using the low-temperature crystalline-water-ice optical constants of \citet{Mastrapa2008,Mastrapa2009}, is shown in Fig.~\ref{fig:epoch3_model_comparison}. The comparison illustrates how the presence of water-ice--amorphous-carbon aggregates in the preferred Epoch~3 architecture facilitates the visibility of the Fresnel structure relative to an architecture in which water ice is associated only with pyroxene.

These results demonstrate that fine-grained aggregate particles containing water ice and refractory materials can strongly suppress the weaker near-infrared water-ice bands while preserving a prominent 3~\textmu{}m absorption feature. Consequently, identifying water ice within the 3~\textmu{}m spectral complex requires consideration of the full band morphology rather than solely the presence or absence of the 1.5 and 2.0~\textmu{}m absorptions. More generally, the morphology of the 3~\textmu{}m absorption complex, including the strength and visibility of the Fresnel reflection feature, reflects not only the phase and temperature of the ice, but also aggregate composition, refractory abundance, and particle size.

\begin{figure*}[ht!]
\plotone{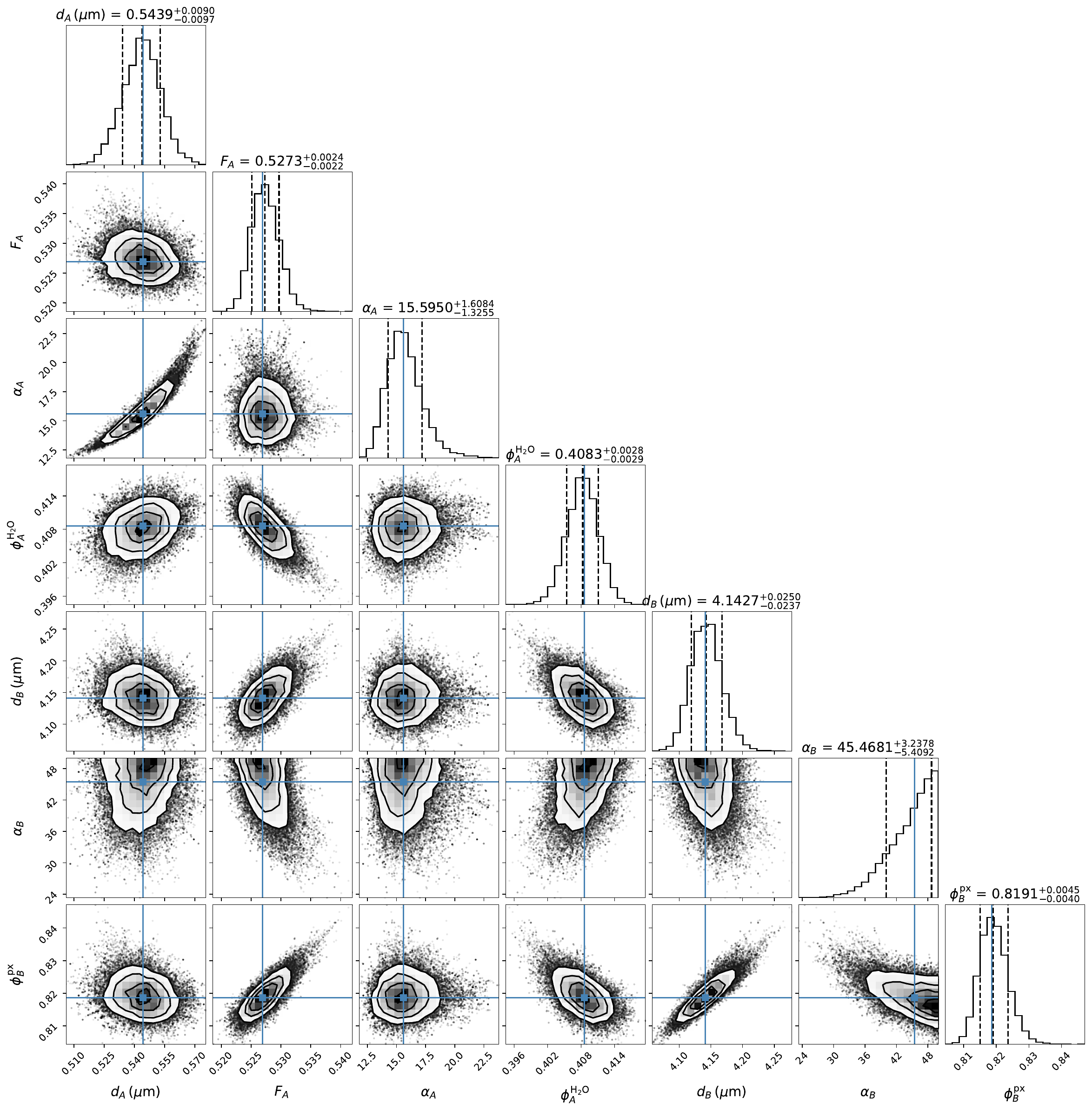}
\caption{
Posterior distributions for the preferred Epoch~1 aggregate model. Blue lines mark the adopted best-fit values (see Table~\ref{tab:epoch_comparison}). Aggregate~A corresponds to the fine water-ice--pyroxene component, whereas Aggregate~B corresponds to the pyroxene--amorphous-carbon continuum component. The strongest covariance occurs between the minimum particle size and particle-size distribution exponent of Aggregate~A, indicating that the spectrum primarily constrains the effective population of fine ice-bearing grains rather than $d_{\min,A}$ and $\alpha_A$ independently. Contours in the two-dimensional histograms represent the 1$\sigma$, 2$\sigma$, and 3$\sigma$ credible regions. Dashed lines in the one-dimensional histograms indicate the 16th, 50th (median), and 84th percentiles. The posterior distribution of $\alpha_B$ is truncated at the adopted upper prior boundary ($\alpha_B = 50$). The preference of the model for such high $\alpha_B$ values indicates a very steep particle-size distribution for Aggregate~B, implying that its optical response is dominated by the smallest particles in the distribution.
\label{fig:mcmc_epoch1}}
\end{figure*}

\begin{figure*}[ht!]
\plotone{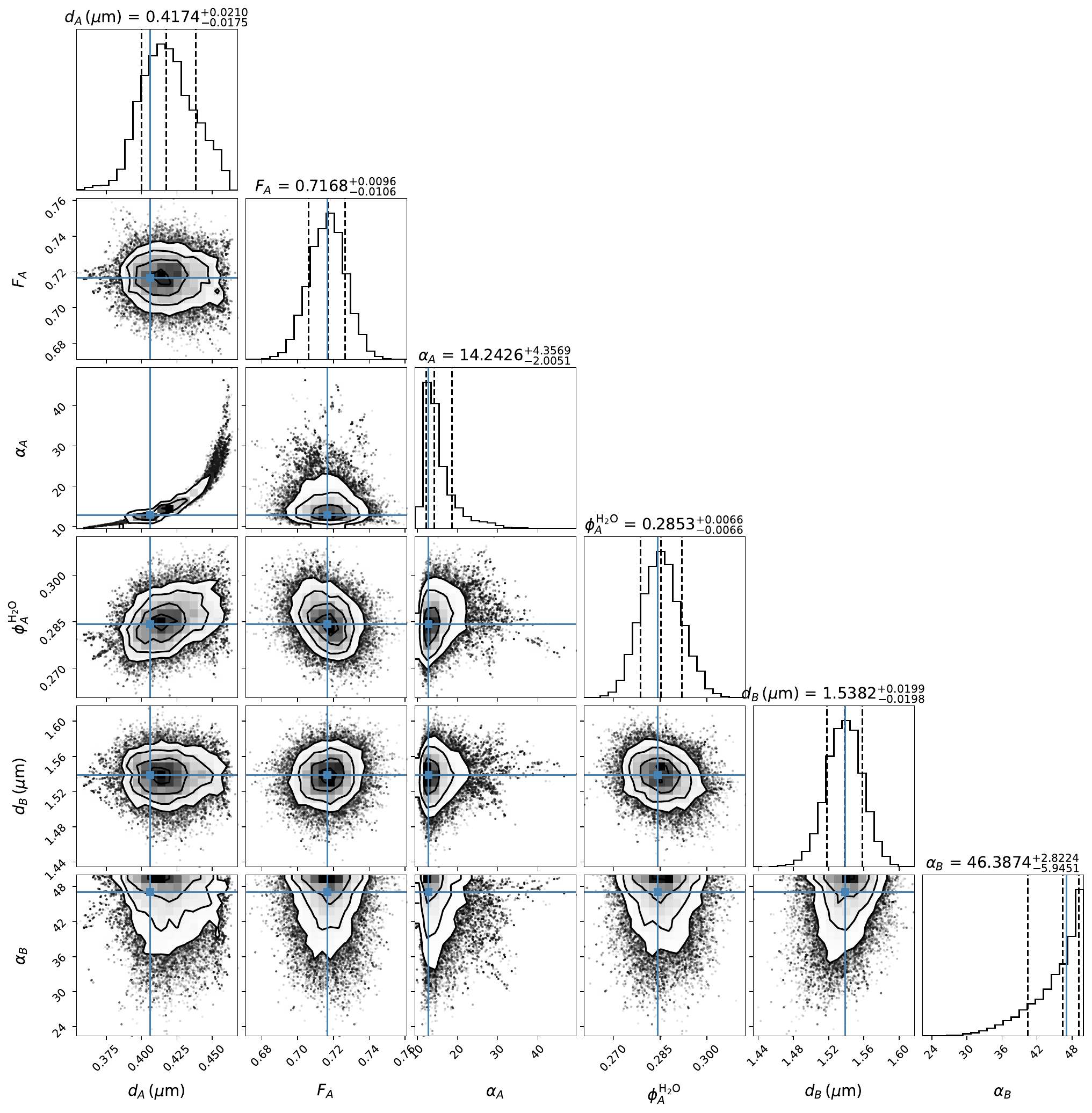}
\caption{
Posterior distributions for the preferred Epoch~3 aggregate model. Blue lines mark the adopted best-fit values (Table~\ref{tab:epoch_comparison}). Aggregate~A corresponds to the fine water-ice--pyroxene component, whereas Aggregate~B corresponds to the water-ice--amorphous-carbon component. The strongest covariance occurs between the minimum particle size and particle-size distribution exponent of Aggregate~A, indicating that the spectrum primarily constrains the effective abundance of fine water-ice--pyroxene aggregates rather than $d_{\min,A}$ and $\alpha_A$ independently. Contours in the two-dimensional histograms represent the 1$\sigma$, 2$\sigma$, and 3$\sigma$ credible regions. Dashed lines in the one-dimensional histograms indicate the 16th, 50th (median), and 84th percentiles. The posterior distribution of $\alpha_B$ is truncated at the adopted upper prior boundary ($\alpha_B = 50$). The preference of the model for such high $\alpha_B$ values indicates a very steep particle-size distribution for Aggregate~B, implying that its optical response is dominated by the smallest particles in the distribution.
\label{fig:mcmc_epoch3}}
\end{figure*}
\section{Possible Contribution of Organic Absorptions}\label{sec:appendix_organics}

The preferred water-ice model does not fully reproduce weak residual structure observed in the Epoch~1 spectrum between 3.3 and 3.6~\textmu{}m (Fig.~\ref{fig:comparison}d). Several of these residuals occur near the wavelengths of absorptions observed on the surface of 67P at 3.38, 3.42, and 3.47~\textmu{}m, which have been attributed to aliphatic CH$_3$ and CH$_2$ stretching modes \citep{Raponi2020}. Although the overall morphology of the long-wavelength wing of the 3~\textmu{}m absorption is broadly similar in the two spectra, the features in 3I are substantially weaker and remain less securely identified. In contrast, we do not identify a distinct 3.10~\textmu{}m absorption analogous to that reported for 67P and interpreted as arising from N--H-bearing species \citep{Raponi2020}. Likewise, any weak absorption near 3.3~\textmu{}m, potentially associated with N--H vibrations or aromatic C--H stretching \citep{Raponi2020}, could be partially obscured by the strong gas emission present in this spectral region. The present data therefore do not permit a unique compositional interpretation, and the contribution of these additional absorptions remains speculative. The lower signal-to-noise ratio of the Epoch~3 spectrum, together with the presence of a Fresnel-like spectral structure
near 3.1~\textmu{}m, precludes a comparable assessment.
\bibliography{Protopapa}{}
\bibliographystyle{aasjournalv7}



\end{document}